\documentclass[twocolumn,showpacs,aps,prl,superscriptaddress,letterpaper]{revtex4}

\usepackage{graphicx}
\usepackage{dcolumn}
\usepackage{amsmath}
\usepackage{epsfig}
\usepackage{multirow}
\usepackage{subfigure}

\long\def\inst#1{\par\nobreak\kern 4pt\nobreak
    {\itshape #1}\par\vskip 10pt plus 3pt minus 3pt}
\RequirePackage{xspace}
\usepackage{relsize}

\newcommand{\BABARPubYear}    {2012}
\newcommand{\BABARPubNumber}  {019}

\newcommand{\SLACPubNumber} {15264}

\newcommand{\LR}{\texttt{L}\xspace}
\newcommand{\HR}{\texttt{H}\xspace}

\input babarsym

\def\cpoddhiggs      {\ensuremath{{A^{0}}}\xspace}

\def\n1Spipi     {\ensuremath{}\xspace}

\def\beq{\begin{equation}}
\def\eeq{\end{equation}}
\def\bea{\begin{eqnarray}}
\def\eea{\end{eqnarray}}
\def\bq{\begin{quote}}
\def\eq{\end{quote}}
\def\bi{\begin{itemize}}
\def\ei{\end{itemize}}
\def\bc{\begin{center}}
\def\ec{\end{center}}

\newcommand{\ie}{{\em i.e.}}

\def\etal{{\em et al.}}

\newcommand{\higgsmass}{\ensuremath{m_{A^0}}\xspace}

\newcommand{\recMass}{\ensuremath{m_\mathrm{recoil}}\xspace}
\newcommand{\missMass}{\ensuremath{m_X^2}\xspace}

\begin{document}

\onecolumngrid
\hbox to \hsize{
\vbox{

\begin{flushleft}
October 20, 2012 \\
\end{flushleft}
\vspace{\baselineskip}
}
\hfill
\vbox{
\begin{flushright}
\babar-PUB-\BABARPubYear/\BABARPubNumber \\
SLAC-PUB-\SLACPubNumber \\
\end{flushright}
}}
\vspace{-\baselineskip}
\twocolumngrid
\title{
\large \bfseries \boldmath
Search for a Low-Mass Scalar Higgs Boson Decaying to a Tau Pair in Single-Photon Decays of $\Upsilon(1S)$
}

%
%
\author{J.~P.~Lees}
\author{V.~Poireau}
\author{V.~Tisserand}
\affiliation{Laboratoire d'Annecy-le-Vieux de Physique des Particules (LAPP), Universit\'e de Savoie, CNRS/IN2P3,  F-74941 Annecy-Le-Vieux, France}
\author{J.~Garra~Tico}
\author{E.~Grauges}
\affiliation{Universitat de Barcelona, Facultat de Fisica, Departament ECM, E-08028 Barcelona, Spain }
\author{A.~Palano$^{ab}$ }
\affiliation{INFN Sezione di Bari$^{a}$; Dipartimento di Fisica, Universit\`a di Bari$^{b}$, I-70126 Bari, Italy }
\author{G.~Eigen}
\author{B.~Stugu}
\affiliation{University of Bergen, Institute of Physics, N-5007 Bergen, Norway }
\author{D.~N.~Brown}
\author{A.~Georges}
\author{L.~T.~Kerth}
\author{Yu.~G.~Kolomensky}
\author{G.~Lynch}
\author{U.~Paudel}
\affiliation{Lawrence Berkeley National Laboratory and University of California, Berkeley, California 94720, USA }
% ...
\collaboration{The \babar\ Collaboration}
\noaffiliation


\begin{abstract}
We search
for a low-mass scalar CP-odd Higgs boson, $A^0$, produced in the
radiative decay of the Upsilon resonance and decaying into a
$\tau^+\tau^-$ pair:
$\Y1S\to\gamma A^0$. 
The production of \Y1S mesons is tagged by $\Y2S\to\pi^+\pi^-\Y1S$
transitions, using a sample of $(98.3\pm0.9)\times 10^6$ \Y2S mesons collected 
by the \babar\ detector. 
We find no evidence for a Higgs boson in the mass range 
$3.5~\mathrm{GeV}\leq\higgsmass\ \leq 9.2$~GeV, and
combine these results with our previous search for the tau
decays of the light Higgs in radiative \Y3S\ decays, setting limits
on the coupling of $A^0$ to the $b\bar{b}$ quarks in the range
$0.09-1.9$. 
Our measurements improve the constraints on the parameters of the
Next-to-Minimal-Supersymmetric Standard Model and similar theories with
low-mass scalar degrees of freedom. 
\end{abstract}


\maketitle


The Higgs boson is a scalar elementary particle predicted by 
the Higgs mechanism~\cite{ref:Higgs} which attempts to explain the
origin of mass of the elementary particles within the Standard Model
(SM)~\cite{ref:SM}. 
Present experimental evidence suggests a Higgs-like state with the mass
of $\approx 126$~GeV~\cite{ref:Higgs_LHC}. 
%
%
However, low-mass Higgs states with masses of
$\mathcal{O}(10~\mathrm{GeV})$ appear in several extensions to the SM
\cite{Dermisek:2006py}, such as
the Next-to-Minimal Supersymmetric Standard Model
(NMSSM)~\cite{ref:NMSSM}. 
The NMSSM adds a singlet
superfield to the Minimal Supersymmetric Standard Model, 
solving the so-called naturalness problem. The NMSSM contains 
two charged Higgs bosons, three neutral \CP-even bosons, and two \CP-odd
bosons. The lightest \CP-odd state, $A^0$, could have a mass below 
the $b\bar{b}$ production threshold, avoiding detection at LEP~\cite{Dermisek:2006py}. 
Such a particle could be produced in radiative $\Upsilon\to\gamma A^0$ 
decays~\cite{Wilczek:1977zn} with a branching fraction as large as
$10^{-4}$ for the narrow states $\Upsilon(nS)$ (where $n\le3$), 
depending on the $A^0$ mass and couplings~\cite{ref:NMSSM}, making it
accessible at B-Factories. Thus, constraining the low-mass NMSSM Higgs
sector is important for understanding the recent LHC
discovery~\cite{LisantiWacker2009}.  

Searches for $A^0$ decays into
$\mu^+\mu^-$~\cite{Higgs2mumu}, $\tau^+\tau^-$~\cite{Higgs2tautau}, 
invisible~\cite{Higgs2invisible}, and hadronic~\cite{Higgs2hadrons}
final states have been performed by \babar, so far 
with null results. 
In particular, limits on the product of branching fractions 
$\BR(\Y3S\to\gamma A^0)\times\BR(A^0\to\tau^+\tau^-)$ have been 
set at $(1.5-16)\times10^{-5}$ in the
mass range $4.03<\higgsmass<10.10$~GeV~\cite{Higgs2tautau}. 
The CLEO Collaboration has set limits on the branching ratio product
$\BR(\Y1S\to\gamma A^0)\times\BR(A^0\to\tau^+\tau^-)$ in the range
$10^{-5}-10^{-4}$ for masses $\higgsmass<9.2$~GeV~\cite{4}. 

This paper describes a search for decays of the $\Upsilon(1S)$
resonance into a photon and a light scalar CP-odd Higgs boson
$A^0$, which then decays into a pair of tau leptons.  The
$\Upsilon(1S)$ resonance is produced from the $\Upsilon(2S)$ resonance
with the emission of two charged pions. The reaction chain is
$e^+ e^- \to \Upsilon(2S) \to \pi^+ \pi^- \Upsilon(1S)$, $\Upsilon(1S)
\to\gamma A^0$, $A^0 \to \tau^+ \tau^-$.   We identify the \Y1S by the 
dipion transition; the production and decay of the Higgs candidates are
identified by the photon and the two 
charged tracks from one-prong decays of the two tau leptons.

This analysis is based on a sample of $(98.3\pm0.9)\times 10^6$ \Y2S\ decays 
collected at the \Y2S resonance with the
\babar\ detector~\cite{detector} at the PEP-II 
asymmetric-energy $e^+e^-$ collider at the SLAC National Accelerator Laboratory.  
This sample corresponds to an integrated luminosity of 14~fb$^{-1}$.  
We also use a sample of 28 fb$^{-1}$ taken at the \Y3S resonance for
studies of the QED (continuum) backgrounds 
and the optimization of the selection of the \Y2S dipion
transition candidates. 
\Y3S\ decays are rejected by the analysis selection criteria
due to their different kinematics distributions compared with the
\Y2S\ decays, and 
therefore the \Y3S\ events form	a pure high-statistics continuum QED sample.
An additional dataset of 1.4~fb$^{-1}$ taken at a center-of-mass
(CM) energy 30~MeV below the \Y2S\ 
mass is used for studies of systematic effects.
We use Monte Carlo (MC) simulated samples of signal and
\Y1S\ background events~\cite{ref:EvtGen,geant} to tune the selection of the
Higgs events. The tau-lepton branching fractions are fixed to the
values of Ref.~\cite{PDBook}.
The \babar\ detector, including the Instrumented Flux Return, the electromagnetic calorimeter, 
and the tracking and particle identification (PID) systems, 
is described in detail elsewhere \cite{detector, LST}.

A signal candidate consists of a photon plus four charged tracks:
$\pi^+ \pi^-$ from the $\Y2S\to\pi^+\pi^-\Y1S$ transition, and the
one-prong decays of the two tau leptons. The event may contain as many
as 19 additional photons with laboratory energy greater than 30~MeV,
mostly from beam-induced backgrounds, but no 
additional charged tracks. Additional signal candidates may be formed
using these extra photons, but a single final candidate per event is
selected, as described below.

We select events where at least one tau lepton decays leptonically, 
resulting in five different combinations of tau lepton daughters: $ee, e\mu, e\pi, \mu\mu, \mu\pi$. Events 
in which both tau leptons decay hadronically suffer from
significantly larger and poorly modeled backgrounds than the leptonic 
channels, and are therefore excluded. The tau lepton daughters are
identified using multivariate discriminants based on the information
from all subdetectors. Typical PID efficiencies are 98\%
($e$), 90\% ($\mu$), and 97\% ($\pi$), while the typical pion
misidentification rates are less than $0.5\%$ ($e$) and 5\% ($\mu$). 
Requirements on the electromagnetic shower shapes
of the primary photon candidates are also imposed to improve the signal purity,
and events with $\pi^{0}$ candidates, formed from pairs of photons
with invariant mass satisfying $100<m_{\gamma\gamma}<160$ MeV and
laboratory energy above 200~MeV, are discarded. 

In order to achieve a balanced selection efficiency that does not
depend strongly on the reconstructed Higgs mass \higgsmass\ (or photon
energy), we optimize the selection in two Higgs mass intervals:
$3.6\le\higgsmass\le8.0$~GeV (\LR\ range) and $8.0<\higgsmass\le9.2$
GeV (\HR\ range). 
The choice of the mass ranges is motivated by the rapidly varying
signal-to-background ratio at low photon energies (which correspond to
higher \higgsmass\ for the signal), and 
differences in kinematics for each photon energy range. 

The masses of the \Y1S\ and \cpoddhiggs\ candidates are calculated
from two primary 
kinematic variables:
\begin{eqnarray}
\recMass^{2} &=& M_{\Y2S}^2 + m_{\pi\pi}^{2} - 2M_{\Y2S}E^{CM}_{\pi\pi}\ ,
\label{eq:Mrecoil}\\
\missMass &=& (P_{e^{+}e^{-}} - P_{\pi\pi} - P_{\gamma})^{2} \ .
\label{eq:Mmiss}
\end{eqnarray}
Here $\recMass$ is the recoil mass of the dipion system (which peaks at the
value of the \Y1S mass for signal), $\missMass$ is the mass recoiling
against the signal photon in the \Y1S\ frame (which peaks near the square of the
expected Higgs mass, $\higgsmass^2$, and is linear in
the photon energy), and
$P$ denotes the four-momentum. 

In order to reject backgrounds,
we train two multilayer perceptron neural 
networks (NN)~\cite{TMVA}: a pion discriminant ($\mathcal{N}_\pi$), which 
describes the kinematics of the process $\Y2S\to \pi^+ \pi^- \Y1S$, and 
a tau discriminant ($\mathcal{N}_\tau$) describing the transition 
$\Y1S\to\gamma A^0,  A^0\to\tau^+\tau^-$. Each NN uses kinematic variables 
only weakly correlated to $\missMass$ or $\recMass$. The two discriminants
are uncorrelated. 
The pion discriminant $\mathcal{N}_\pi$ combines nine kinematic
variables specific to the dipion
system~\cite{BAD2009prl,Higgs2invisible}. The discriminant 
$\mathcal{N}_\tau$ is a combination of 14 variables: the missing energy
and the cosine of the polar angle of the missing momentum in the
event; the extra
calorimeter energy in the lab frame 
and the energy of the second most energetic photon in the CM frame; the
net transverse momentum of the reconstructed signal candidate particles;
the acoplanarity 
of the photon relative to the plane formed by the two tau decay prongs;
the momentum and polar angle in the CM frame of the most energetic tau
decay prong; the invariant mass, vertex probability, and the distance
between the vertex and the $e^+e^-$ interaction region of the two tau
decay prongs; the angle between the signal photon and the most energetic
tau decay prong in the event. 
The discriminants are trained using signal MC events 
$\Y1S \to \gamma A^0$ in the range $4.0 \leq \higgsmass \leq 9.2$
GeV. The background samples for training are taken from the continuum
sample for the pion discriminant and from the simulated generic
\Y1S\ decays for the tau discriminant. 

Each NN outputs a value $\mathcal{N}$ close to $+1$ for signal and
to $-1$ for background.  
Based on the NN outputs, 
the selection criteria for $\mathcal{N}_\pi$ and $\mathcal{N}_\tau$
are chosen to optimize $\varepsilon/(1.5+\sqrt{B})$~\cite{Punzi},  
where $\varepsilon$ is the signal efficiency, and $B$ is the expected
background yield. 
We accept signal candidates if
$\mathcal{N}_{\tau}$ is above a threshold value chosen individually
for each final state and mass range. The thresholds for $\mathcal{N}_\pi$ are
the same for all final states, but different for the two mass ranges. 
The typical signal efficiency and background rejection 
estimated from the corresponding MC samples are listed in
Table~\ref{tab:NNeff}.
\begin{table}[h]
\caption{Typical selection efficiencies (SE) and background rejection
  factors (BR) for
  the two neural network discriminants  $\mathcal{N}_\pi$ and
  $\mathcal{N}_\tau$ in the \LR ($3.6\le\higgsmass\le8.0$~GeV) and \HR
  ($8.0<\higgsmass\le9.2$~GeV) ranges. The SE and BR factors are
  relative to the preselection that requires 4 tracks and a real
  photon in the final state, 
  and are averaged over each mass range.}  
\label{tab:NNeff}
\begin{center}
\begin{tabular}{l|l|c|c}
\hline
\hline
Mass Range & Discriminant &{SE (\%)} & {BR (\%)}\\
\hline
\hline
\LR & $\mathcal{N}_\pi$ & 76 & 99 \\
\HR & $\mathcal{N}_\pi$ & 72 & 99 \\
\LR & $\mathcal{N}_\tau$ & 80 & 97  \\
\HR & $\mathcal{N}_\tau$ & 30 & 99 \\
\hline
\end{tabular}
\end{center}
\end{table}

Due to reconstruction ambiguities, in particular extra photon
candidates in the event and a large $\mu$-as-$\pi$ misidentification
rate, a fraction of signal and background 
events have more than a single reconstructed candidate. 
The multiplicity of candidates per event is on average $1.8$ for 
the simulated signal samples, $1.6$ for the generic \Y2S\ decays, $1.3$ 
for the continuum sample, and $1.5$ for the data. 
We select a single
candidate based on (1) the highest value of $\mathcal{N}_\pi$, then,
if multiple candidates still remain, (2) the highest value of
$\mathcal{N}_\tau$, and finally, (3) the tau decay final state with the
highest signal/background ratio. 

\begin{figure}[h]
\begin{center}
	  \includegraphics [width=0.5\textwidth]{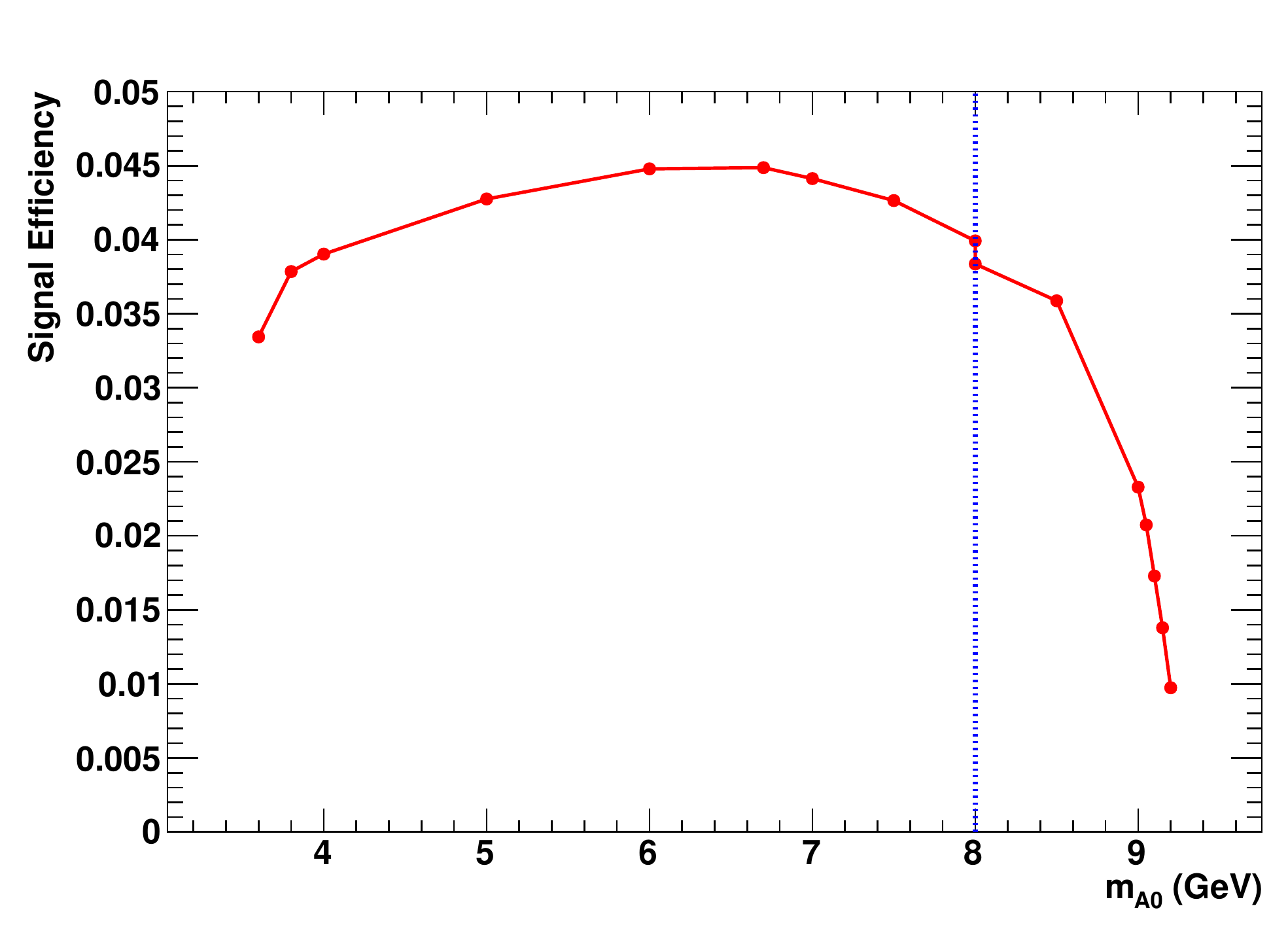}
\caption{Signal efficiency as a function of \higgsmass\ in the \LR\ and
  \HR\ mass ranges (separated by the vertical blue line).}
\label{fig:effSig}
\end{center}
\end{figure}

We further 
suppress the continuum background by applying a cut on the
mass recoiling against the dipion system $\recMass$: 
\begin{equation}
|\recMass-\langle \recMass\rangle|<10~\mathrm{MeV}\ ,
\end{equation}
where $\langle \recMass\rangle$ is the expected location of
the $\Upsilon(1S)$ peak, determined by the mass difference between
the \Y2S\ and \Y1S mesons~\cite{Y2S-Y1Sdiff}.  The final signal selection
efficiency varies between $1\%$ and 
$4.5\%$ (Fig.~\ref{fig:effSig}), and is lowest at the highest masses
(lowest photon energy).  


We extract the yield of signal events as a function of \higgsmass
in the interval $3.6 \leq \higgsmass \leq 9.2$ GeV  
by performing a series of maximum likelihood fits in steps of 
\higgsmass. 
We perform one-dimensional unbinned extended maximum likelihood (ML)
fits to the distribution of \missMass\ 
in the intervals $12 \leq \missMass \leq 72$ GeV$^2$ (\LR\ range)
and  $49 \leq \missMass \leq 89$ GeV$^2$ (\HR\ range).The fit intervals
overlap to provide sufficient sidebands for all values of \higgsmass. 
The likelihood contains contributions from signal,
which is expected to peak near the Higgs mass squared, and from
the smooth background function, arising from continuum and radiative
leptonic \Y1S backgrounds.  
We search for the $A^0$ in varying mass steps that correspond to
approximately half of the expected resolution on \higgsmass, as
described below. A total of 201 mass points are sampled. 

We use signal MC samples
$\Y2S\to\pi^+\pi^-\Y1S,\,\Y1S\to\gamma\cpoddhiggs$  generated at 
nine (seven) different values of \higgsmass in the \LR\ (\HR) mass
range to determine the signal probability density functions (PDFs) in
\missMass\ and selection 
efficiencies. We interpolate these distributions and efficiencies
between fixed \higgsmass\ points, correcting 
for known small differences between data and MC
simulations.  
The signal PDFs only include events in
which the simulated signal 
photon, $\pi^+\pi^-$ tracks, and the tracks from the decay of the tau
leptons are correctly 
reconstructed. The signal efficiency, however, includes
contributions from events in which one of
the final-state particles may be misreconstructed. 
The signal PDF is modeled as a Crystal Ball function
\cite{ref:CBshape}. 
We parameterize the background PDFs as
\begin{eqnarray*}
f(z) &=& \left(\mathrm{Erf}\left[\alpha(z-z_{\min})\right]+1\right)C_3(z)\ (\mathrm{\LR\ range})\ ;\\
f(z) &=& \left(1-\frac{z}{z_{\max}}\right)^\beta\exp[C_2(z)]\ (\mathrm{\HR\ range})
\ ,
\end{eqnarray*}
where $z\equiv\missMass$, $C_n(z)$ is an $n$th-order Chebyshev
polynomial (with different parameters for each mass range), $\alpha$
and $\beta$ are threshold parameters, and 
$z_{\min}=12.5$~GeV$^2$ and 
$z_{\max}=88.9$~GeV$^{2}$ are determined by the kinematic end points of the
photon energy spectrum in the luminosity-weighted mixture of simulated generic \Y1S\ decays
and continuum sample. A common background PDF
describes all five decay channels adequately well.
Parameters of the Chebyshev
polynomials $C_n(z)$ and the threshold parameters $\alpha$ and $\beta$
are determined from a fit to the data 
distribution of $\missMass$ with the signal yield fixed to zero. This
accounts for uncertainties in the modeling of the radiative
\Y1S\ decays and additional backgrounds that may not be well described
by the continuum sample, such as two-photon events with low photon
energy (high \missMass). 
For each \higgsmass\ hypothesis, we determine two parameters: the
background yield 
$N_\mathrm{bkg}$ and
the signal yield $N_\mathrm{sig}$. In the \HR\ range, we also allow the
two coefficients of $C_2(z)$ and the parameter $\beta$ to vary in the fit. 
The fit is performed simultaneously over the distributions in each
$\tau^+\tau^-$ decay channel, taking advantage of the difference in
the signal-to-background ratios over the decay channels. The fraction
of events in each channel is fixed from MC samples for signal, and
from the luminosity-weighted mixture of simulated generic \Y1S\ decays
and continuum sample for the background.


Each fixed nuisance parameter in the fit is varied according to its
uncertainty; correlations between parameters are taken into
account. 
The systematic uncertainties for this measurement can be divided into
two categories:
\begin{itemize}
\item {\em Additive errors}: uncertainties on the event yield, which do not
  scale with the number of reconstructed signal events. These include
  uncertainties of the parameters fixed in the fit (PDF shape
  parameters for signal and backgrounds) and a small bias in the
  ML fit. These
  uncertainties reduce the significance of any observed
  signal~\cite{ref:Hooberman}.  
\item {\em Multiplicative errors}: uncertainties that scale with the number
  of reconstructed signal events. These include uncertainties on the
  reconstruction efficiency, the ML fit bias which scales with the true number
  of signal events, the uncertainty in the number of produced
  \Y2S\ mesons, and the 
  uncertainty in the branching fraction of $\Y2S\to\pi^+\pi^-\Y1S$. 
\end{itemize}

We compute the average bias of the ML fit for a set of 
generated $N_\mathrm{sig}$ values using a large ensemble of simulated
pseudo-experiments. In each pseudo-experiment, the signal events are
fully simulated, and the background events are sampled from their
PDFs. We determine the fit bias that is independent of
$N_\mathrm{sig}$ and is part of the additive uncertainties, as well
as the bias which scales linearly with $N_\mathrm{sig}$, and  
can be thought of as a ``fit inefficiency'', \ie\ a relative
correction to the signal reconstruction efficiency. The bias arises
from imperfections in modeling of the signal PDFs, from events in which
signal candidates are misreconstructed, and from low-statistics properties
of the ML estimators. 
We see that a
correction of $3.1\pm1.1\%$ (\LR\ range) and $7.6\pm1.4\%$
(\HR\ range) has to be applied. The additive parts of the fit bias are
small. 

The signal efficiencies determined in MC simulations are corrected by
several multiplicative effects: 
\begin{itemize}
\item {\em Tracking and dipion selection efficiency}. These
  corrections and their uncertainties have been
  determined~\cite{Higgs2invisible} 
  using a clean sample of of four-track final states from decays
  $\Y2S\to\pi^+\pi^-\Y1S,\ \Y1S\to\mu^+\mu^-$. The data/MC 
  ratio of $0.97\pm0.02$ includes the uncertainties due
  to the number of produced \Y2S\ events, dipion branching ratio
  $\Y2S\to\pi^+\pi^-\Y1S$, dipion reconstruction efficiency,
  the efficiency of reconstructing two additional (energetic) charged
  tracks, trigger uncertainties, and the selection efficiency for 
  the pion discriminant $\mathcal{N}_\pi$. 
  The uncertainty is dominated by the error on
  $\Y1S\to\mu^+\mu^-$ branching ratio~\cite{PDBook}. 

\item {\em Photon selection efficiency}. A correction of
  $0.967\pm0.017$  is determined from a high-statistics
  $e^+e^-\to\gamma\gamma$ sample in which one of the photons converts
  in the inner detector material to produce a detectable $e^+e^-$
  pair~\cite{Higgs2invisible}. 

\item {\em Neural Network selection efficiency}. We evaluate the systematic
  uncertainty due to possible data/MC differences in the distributions
  of the NN discriminant $\mathcal{N}_\tau$ using an inclusive background
  sample. We select signal-like events that
  pass the requirement 
  $\mathcal{N}_\tau>0$ and compute
  the ratio of partial selection efficiencies for the actual
  $\mathcal{N}_\tau$ thresholds for the data and the
  background. 
These ratios are $1.038\pm0.013$ (\LR\ range) and $0.991\pm0.014$
(\HR\ range). 
\end{itemize}

The total correction to the efficiency is a product of all corrections
discussed above:
\begin{eqnarray*}
\varepsilon_\mathrm{data}/\varepsilon_\mathrm{MC}
&=0.943\pm0.031\ (\mathrm{\LR\ range})\ ; \\
\varepsilon_\mathrm{data}/\varepsilon_\mathrm{MC}
&=0.859\pm0.033\ (\mathrm{\HR\ range})\ .
\end{eqnarray*}

\begin{figure}[h]
\begin{center}
\subfigure[] {
	 \includegraphics [width=0.5\textwidth]{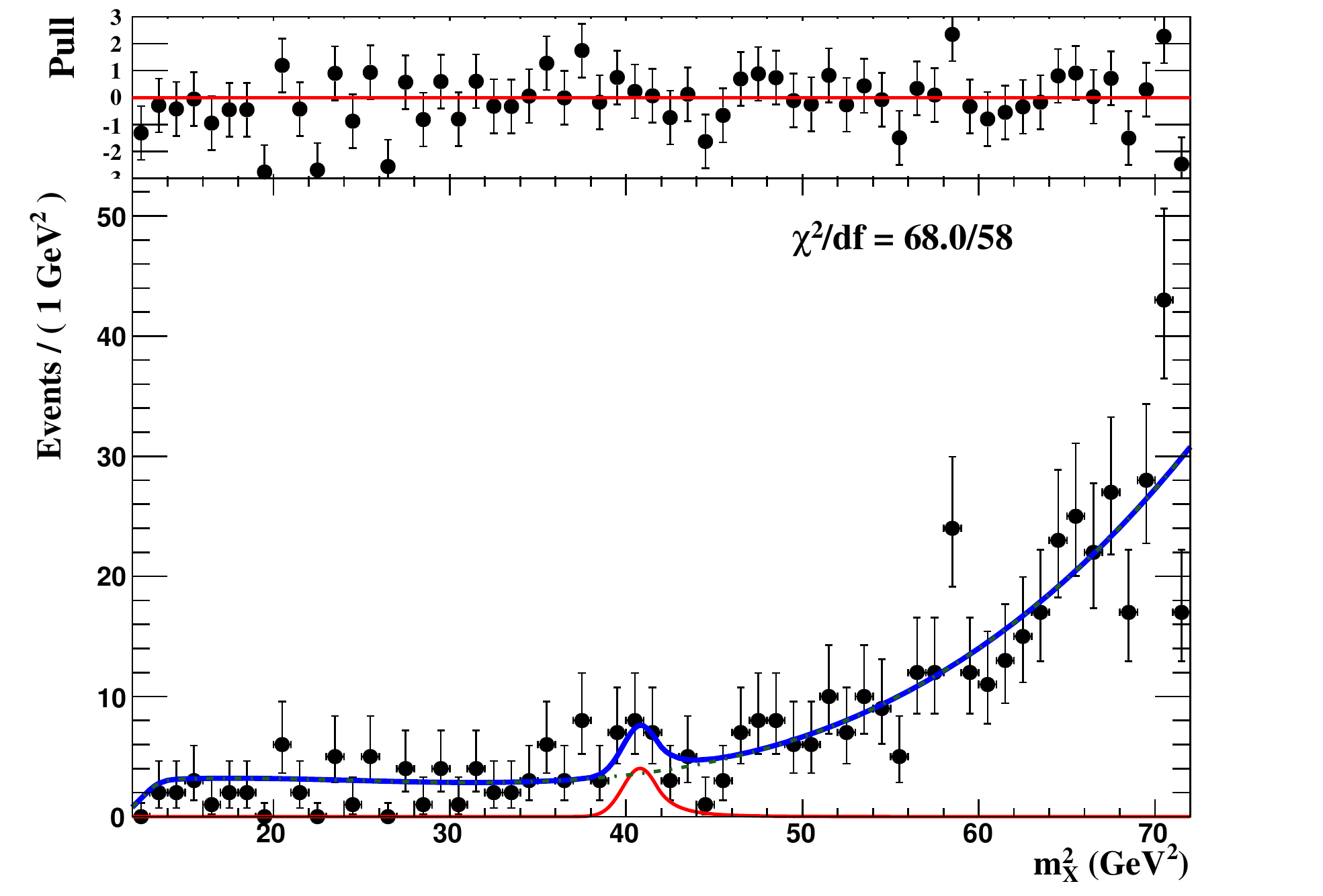}
\label{fig:}
}
\subfigure[] {
	 \includegraphics [width=0.5\textwidth]{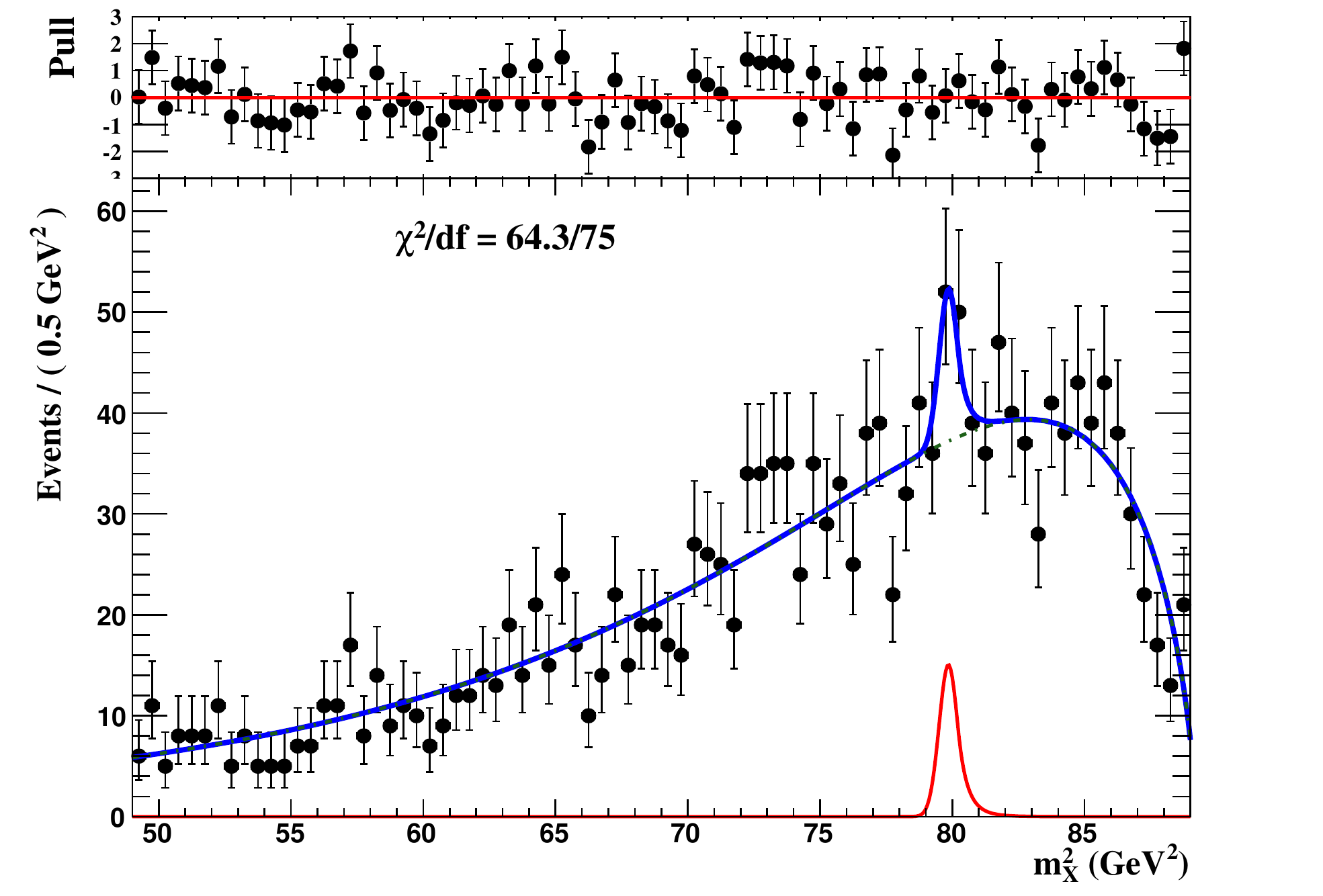}
\label{fig:proj_}
}
\caption{Fits to the \missMass\ distributions in (a) \LR and (b) \HR ranges
  for the two particular \higgsmass points that return the largest
  upward fluctuations: (a) $\higgsmass=6.36$~GeV and (b) $\higgsmass=8.93$~GeV. 
 The red solid line shows the signal PDF,
  the green dot-dashed line is the background contribution, and the blue
  solid line shows the total PDF. The top plots show the fit residuals
normalized by the error (pulls). The signal peaks corresponds to
a statistical significance of (a) $2.7\sigma$ and (b) $3.0\sigma$.}
\label{fig:proj}
\end{center}
\end{figure}

We compute the statistical
significance of a particular fit centered at \higgsmass\ as 
$\mathcal{S}=\sqrt{2\log(\mathcal{L_{\max}}/\mathcal{L}_0)}$, where 
$\mathcal{L_{\max}}$ is the maximum likelihood value for a fit with a
free signal yield, and
$\mathcal{L}_0$ is the value of the likelihood for
$N_\mathrm{sig}=0$. 
Including additive systematic uncertainties, the most significant
upward fluctuations  
occur at $\higgsmass=6.36$~GeV with $\mathcal{S}=2.7\sigma$
(Fig.~\ref{fig:proj}a) and 
$\higgsmass=8.93$~GeV with $\mathcal{S}=3.0\sigma$
(Fig.~\ref{fig:proj}b). 
Fluctuations of $+3\sigma$ or higher occur in $7.5\%$ of
pseudo-experiments that simulate a scan of 201 mass points with an
average correlation of $94.5\%$ between adjacent points, as observed
in our dataset. 
Therefore, we
conclude that no significant \cpoddhiggs\ signal is found. 

\begin{figure}[h]
\begin{center}
	  \includegraphics [width=0.5\textwidth]{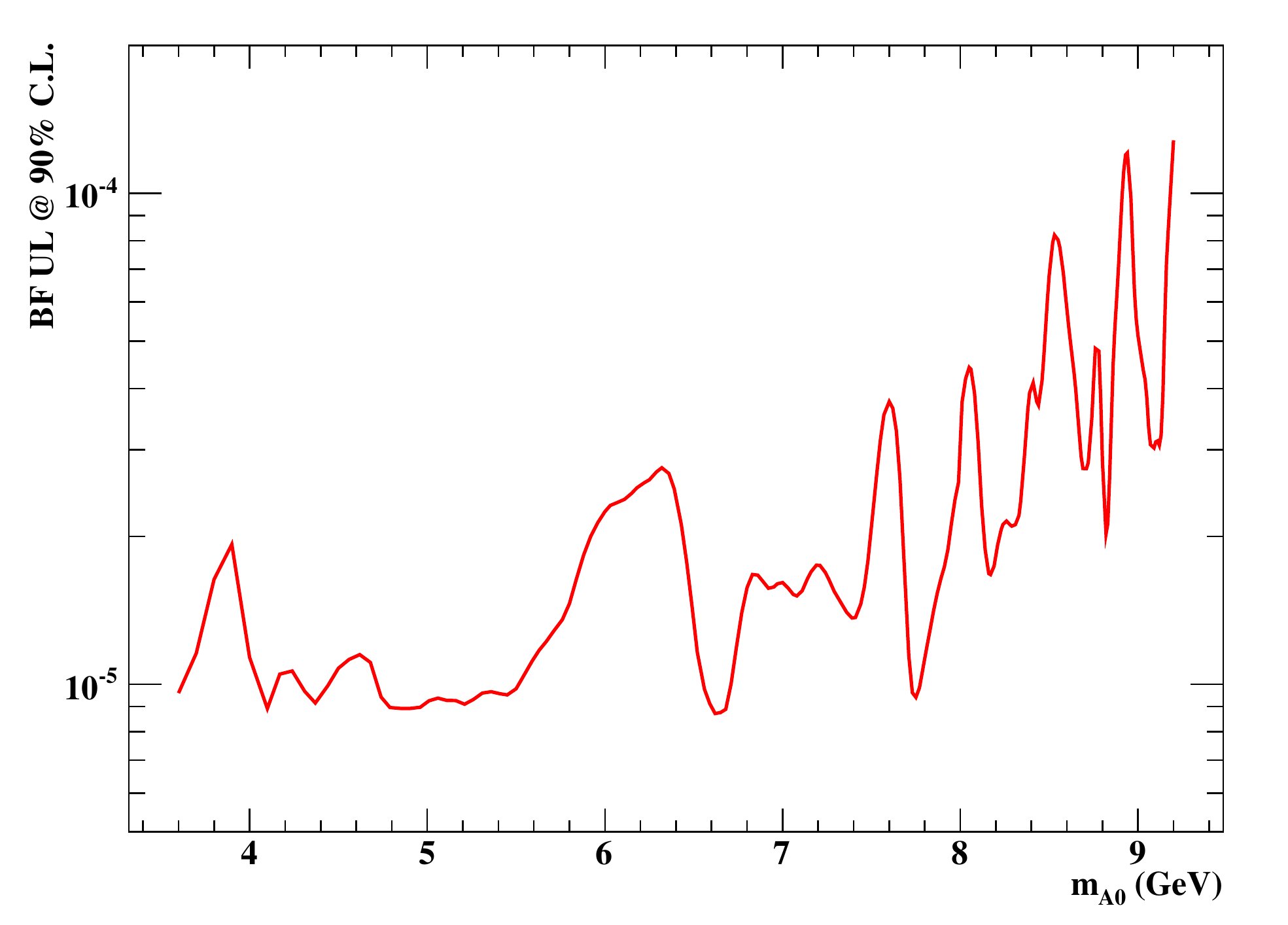}
\caption{90\% C.L. upper limits for $\BR(\Y1S\to\gamma A^0)\times\BR(A^0\to\tau^+\tau^-)$.}
\label{fig:BF}
\end{center}
\end{figure}

Since we do not observe a significant excess of events above the background,
we set 90\% confidence level (C.L.) Bayesian upper limits
on the product $\BR(\Y1S\to\gamma A^0)\times\BR(A^0\to\tau^+\tau^-)$,
computed with a uniform prior for $N_\mathrm{sig}>0$. 
The limits are shown in Fig.~\ref{fig:BF}. Systematic uncertainties on
$N_\mathrm{sig}$ and 
$\varepsilon_\mathrm{data}$  are included assuming their likelihood
profiles are Gaussian~\cite{ref:Hooberman}. 

We combine our results with the previous limits on the branching
ratios $\BR(\Y3S\to\gamma
A^0)\times\BR(A^0\to\tau^+\tau^-)$~\cite{Higgs2tautau} to set a limit
on the Yukawa coupling $g_b^2$ of the $b$-quark to the $A^0$. 
The branching fractions $\BR(\Upsilon(nS)\to\gamma\cpoddhiggs)$ are
related to $g_b$ through~\cite{Wilczek:1977zn,ManganoNason,Nason86}: 
\beq
\frac{\BR(\Upsilon(nS)\to\gamma\cpoddhiggs)}{\BR(\Upsilon(nS)\to l^+l^-)} = 
\frac{g_b^2 G_Fm_b^2}{\sqrt{2}\pi\alpha}\mathcal{F}_{QCD}\left(1-\frac{m^2_{\cpoddhiggs}}{m^2_{\Upsilon(nS)}}\right)\ ,
\label{eq:BFRatio}
\eeq
where $l\equiv e$ or $\mu$, $\alpha$ is the fine structure constant
computed at the scale $m_{\Upsilon(nS)}$, $G_F$ is the Fermi
constant, and   
$\mathcal{F}_{QCD}$ includes
the \higgsmass-dependent QCD and
relativistic corrections to
$\BR(\Upsilon(nS)\to\gamma\cpoddhiggs)$~\cite{Nason86} and  
the leptonic width of $\Upsilon(nS)$~\cite{Barbieri75}. To first order
in $\alpha_S$, the corrections range from 0 to 30\%~\cite{Nason86} and
may have large uncertainties~\cite{Beneke1997}. 

\begin{figure}[h]
\begin{center}
	  \includegraphics [width=0.5\textwidth]{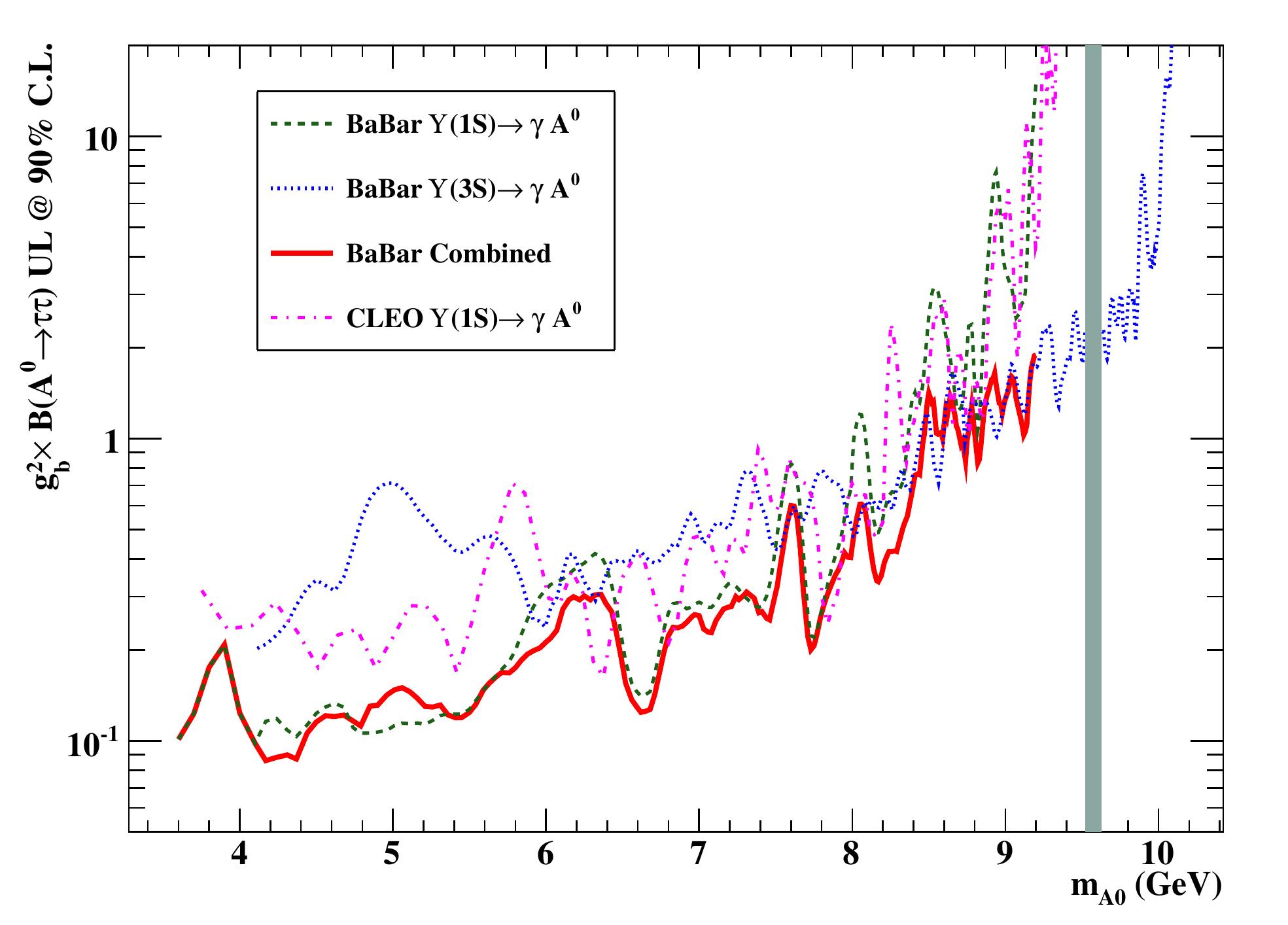}
\caption{90\% C.L. upper limits for Yukawa coupling
  $g_b^2\times\BR(A^0\to\tau^+\tau^-)$. Shown are 
  combined \babar\ results (red solid line),
  results from this
  analysis only (dashed green line), the previous
  \babar\ measurement~\cite{Higgs2tautau} (dotted blue line), and
  results from the CLEO experiment~\cite{4} (dot-dashed black
  line). The shaded vertical bar shows the region around $\chi_b(2P)$
  mass excluded from Ref.~\cite{Higgs2tautau}.}
\label{fig:coupling}
\end{center}
\end{figure}

We combine our data on $\Y1S\to\gamma A^0$ with the \babar\ results of
Ref.~\cite{Higgs2tautau} using the full likelihood functions for
$g_b^2$ at each \higgsmass\ point from this analysis, and a Gaussian
approximation for the $g_b^2$ likelihood from
Ref.~\cite{Higgs2tautau}. The combined upper limits on the product
$g_b^2\times\BR(A^0\to\tau^+\tau^-)$ as a function of \higgsmass\ are
shown in Fig.~\ref{fig:coupling}. They rule out much of the parameter
space preferred by NMSSM $g_b = \tan\beta\cos\theta_A > 1$, where
$\tan\beta$ is the ratio of the Higgs vacuum expectation values and
$\cos\theta_A$ is the fraction of the non-singlet component in the
CP-odd $A^0$~\cite{ref:NMSSM}.

In summary, we find no evidence for the radiative decays
$\Y1S\to\gamma A^0$ in which $A^0$ decays into a pair of tau leptons,
and we set 90\% C.L. upper limits 
on the product of branching fractions $\BR(\Y1S\to\gamma
A^0)\times\BR(A^0\to\tau^+\tau^-)$ 
in the range $(0.9-13)\times10^{-5}$ for
$3.6 \le\higgsmass\le9.2\,\gev$. We also set 90\% C.L. upper limits  
on the product $g_b^2\times\BR(A^0\to\tau^+\tau^-)$ in the range
$0.09-1.9$ for $\higgsmass\le9.2\,\gev$. Our limits place significant
constraints on NMSSM parameter space.

We are grateful for the excellent luminosity and machine conditions
provided by our \pep2\ colleagues, 
and for the substantial dedicated effort from
the computing organizations that support \babar.
The collaborating institutions wish to thank 
SLAC for its support and kind hospitality. 
This work is supported by
DOE
and NSF (USA),
NSERC (Canada),
CEA and
CNRS-IN2P3
(France),
BMBF and DFG
(Germany),
INFN (Italy),
FOM (The Netherlands),
NFR (Norway),
MES (Russia),
MICIIN (Spain),
STFC (United Kingdom). 
Individuals have received support from the
Marie Curie EIF (European Union),
the A.~P.~Sloan Foundation (USA)
and the Binational Science Foundation (USA-Israel).


\vfill

\end{document}